\documentstyle[amssymb,prl,aps,epsf]{revtex}

\twocolumn 
\tighten 

\begin{document}
\draft

\twocolumn[ 
\hsize\textwidth\columnwidth\hsize\csname@twocolumnfalse\endcsname 
\title{Ginzburg-Landau theory of an RVB superconductor} 
\author{V. N. Muthukumar$^a$ and Z. Y. Weng$^b$} 
\address{$^a$ Department of Physics, Princeton University,
Princeton, NJ 08544\\
$^b$ Center for Advanced Study, Tsinghua University,
Beijing 100084, China\\
}
\maketitle 
\begin{abstract} 
We present a Ginzburg-Landau formulation of the {\em bosonic} 
resonating-valence-bond (RVB) theory of superconductivity. The 
superconducting order parameter is characterized by phase vortices that 
describe {\em spinon} excitations and the transition to the superconducting
state occurs when such phase vortices (un)bind. We show that the boson RVB 
theory always leads to $hc/2e$ flux quanta, and that the presence 
of a trapped spin-1/2 moment inside a vortex core gives rise to observable 
consequences for the low temperature field-dependent specific heat.  We also 
show that the cores of magnetic fluxoids exhibit enhanced antiferromagnetic correlations. 

\end{abstract} 
] \narrowtext 


Following the original proposal of Anderson \cite{pwa_87}, there is, by now,
a considerable body of literature devoted to the study of the
resonating-valence-bond (RVB) theory. Much of this is based on the so called 
$t-J$ model, which is the simplest possible model describing electrons
(holes) moving in an antiferromagnetic background \cite{zhangrice_88}. The
RVB theories postulate that such a system is best described by collective
spin and charge degrees of freedom (usually called spinons and holons in the
literature), rather than a quasiparticle theory of interacting electrons.
While there is no satisfactory proof that the quasiparticle theory fails in
the $t-J$ or related models, there are some indications for such a failure,
both from theoretical and numerical studies. Phenomenology of the cuprate
superconductors based on these ideas has also been fairly successful.
Motivated by these considerations, we present, in this paper, an effective
theory of a superconducting RVB state. The theory leads to several
interesting experimental consequences, and is very different from the
conventional BCS theory of superconductivity. Since the theory makes
definite predictions, its veracity can be tested easily.

The theory we present, is based on a bosonic description of the $t-J$ model.
We choose this description since it accounts very well for the short-range
antiferromagnetic (AF) correlations that play an important role in the
cuprate superconductors. Further, in the limit of zero doping, the $t-J$
model reduces to the Heisenberg model and the bosonic RVB theory in this
limit (Schwinger-boson mean field theory) provides an excellent description
of the AF long-range ordered ground state, and the excited states \cite
{aa_88}, \cite{yoshioka_89}. In this limit, the bosonic RVB state can also
be related to the variational wave function of Liang, Doucot and Anderson 
\cite{lda_88}, as shown by Chen \cite{chen_92}. Away from half filling, the
theory describes {\em bosonic} charge (holon) and spin (spinon) degrees of
freedom. In the bosonic theory of the RVB state \cite{zy_99}, the electron
operator is expressed in terms of the bosonic holon and spinon operators and%
{\em \ a topological (vortex) phase operator} as 
\begin{equation}
c_{i\sigma }=h_i^{\dagger }b_{i\sigma }e^{i\hat{\Theta}_{i\sigma }}~.
\label{electron}
\end{equation}
It satisfies usual fermionic anticommutation relations. The phase operator $%
\hat{\Theta}_{i\sigma }$ is the most important ingredient of the theory. It
arises because the doped holes move in a spin background (not necessarily
ordered) with AF correlations, and reflects the nonlocal effects of adding a
hole to a doped Mott AF insulator. In what follows, we shall be concerned
with the holon and spinon degrees of freedom, and therefore, it is natural
to ask what role does the phase operator play in the description of these
degrees of freedom. As shown in \cite{zy_99}, rewriting the $t-J$ model in
terms of the bosonic representation (\ref{electron}) leads to the emergence
of two link fields, $A^s$ and $A^h$. The link field $A^s$ is coupled to the 
{\em holon} degrees of freedom and describes fictitious fluxoids bound to 
{\em spinons} satisfying $\sum_cA_{ij}^s=\pm \pi \sum_{l\epsilon c}\sigma
n_{l\sigma }^b$, for an arbitrary closed path $c$. Here, $n_\sigma ^b$
denotes the spinon number operator. Similarly, the link field $A_{ij}^h$ is
coupled to the {\em spinon} degrees of freedom and describes fictitious
fluxoids bound to {\em holons} satisfying $\sum_cA_{ij}^h=\pm \pi
\sum_{l\epsilon c}n_l^h$, where $n^h$ is the holon number operator. Thus, we
are led to the following physical picture. The motion of holes in an AF
background leads to nonlocal correlations between the charge and spin
degrees of freedom, that are described by the link fields. The holons feel
the presence of the spinons as vortices (quantized flux tubes) and {\em vice
versa }(see Fig. 1). Inasmuch the holons and spinons perceive their mutual
presence through the $\pi $ flux quanta, this theory can also be thought of
as a $\pi $ flux theory, albeit one where the $\pi $ fluxoids are {\em bound
to the constituent particles}.

The effective Hamiltonian is given by $H_{{\rm eff}}=H_h+H_s$, where $H_h$
and $H_s$ are the holon and spinon Hamiltonians respectively. Let us begin
with the holon Hamiltonian, $H_h$. For convenience, we work with a continuum
model and write 
\begin{equation}
H_h\approx {\frac 1{2m_h}}\int d^2{\bf r}~h^{\dagger }({\bf r})\left(
-i\nabla -{\bf A}^e-{\bf A}^s\right) ^2h({\bf r})~,  \label{hholon}
\end{equation}
where $m_h$ $\simeq (2t_ha^2)^{-1}$ ($t_h\sim t)$ is the effective mass of
the holon, and ${\bf A}^e$, the vector potential of the external
electromagnetic field. Note that the holon Hamiltonian (\ref{hholon}) is
coupled to the spinons through the term, ${\bf A}^s({\bf r})$. This term is
the continuum version of the link field $A_{ij}^s$ and is given by 
\begin{equation}
{\bf A}^s({\bf r})={\frac 12}\int d^2{\bf r}^{\prime }{\frac{\hat{{\bf z}}%
\times ({\bf r}-{\bf r}^{\prime })}{|{\bf r}-{\bf r}^{\prime }|^2}}\left[
n_{\uparrow }^b({\bf r}^{\prime })-n_{\downarrow }^b({\bf r}^{\prime
})\right] ~.  \label{As}
\end{equation}
The spinon Hamiltonian $H_s$ is given by 
\begin{equation}
H_s\approx -{\frac J2}\sum_{<ij>\sigma }\left[ \Delta _{ij}^s\left(
e^{i\sigma A_{ij}^h}\right) b_{i\sigma }^{\dagger }b_{j-\sigma }^{\dagger }+%
{\rm h.c.}\right] ~,  \label{hspinon}
\end{equation}
where the spinon pairing (RVB) order parameter is defined by $\Delta
_{ij}^s=\sum_\sigma <e^{-i\sigma A_{ij}^h}b_{i\sigma }b_{j-\sigma }>$. Note
again, the presence of the {\em holon} link variables in the definition of
the {\em spinon} pairing (RVB) order parameter. For zero doping, the order
parameter is identical to that of the Schwinger-boson mean field theory. As
shown in \cite{zy_99}, the Hamiltonian (\ref{hspinon}) can be solved for a
fixed hole concentration, within a self-consistent mean field theory. The
bosonic spinon excitations are gapped, and the gap vanishes as the hole
doping decreases to zero. The short-range AF correlations are determined by
the RVB order parameter $\Delta _{ij}^s$.

When the holons (Bose) condense, the system should become superconducting,
since the spinons are already paired. However, the presence of the phase
field $\hat{\Theta}_{i\sigma }$ in (\ref{electron}) leads to interesting
consequences. To see this, let us first write down the superconducting order
parameter in terms of the decomposition (\ref{electron}). Assuming singlet
pairing of electrons on nearest neighbor sites, we get (in the continuum
limit), 
\begin{equation}
\Delta _{\widehat{\eta }}({\bf r})=\Delta _{\widehat{\eta }}^0\;e^{i\Phi ^s(%
{\bf r})}~,  \label{sc}
\end{equation}
where $\Delta _{\widehat{\eta }}^0\;=f_{\widehat{\eta }}\Delta ^s[\psi
_h^{*}]^2$ ( $f_{\widehat{\eta }}=\pm 1$ for $\widehat{\eta }=\widehat{x},%
\widehat{y}$)$,$ $\psi _h^{*}({\bf r})=<h^{\dagger }({\bf r})>$ denotes the
Bose condensate of the holons, and $\Phi ^s({\bf r})$ arising from the phase
field $\hat{\Theta}_{i\sigma }$ in (\ref{electron}) is given by 
\begin{equation}
\Phi ^s({\bf r})=\int d^2{\bf r}^{\prime }~{\rm Im~ln}\left[ z-z^{\prime
}\right] ~\left( n_{\uparrow }^b({\bf r}^{\prime })-n_{\downarrow }^b({\bf r}%
^{\prime })\right) ~,  \label{phis}
\end{equation}
where $z=x+iy$. From (\ref{sc}) and (\ref{phis}), it is clear that $\Phi ^s$
describes phase vortices centered around the spinons: $\Phi ^s\rightarrow
\Phi ^s\pm 2\pi $, if the coordinate ${\bf r}$ winds around a spinon
continuously in space. Evidently, $\Phi ^s$ is related to ${\bf A}^s$ ($%
\nabla \Phi ^s=2{\bf A}^s$) and both describe the vortex effect related to
the spinon excitations. From (\ref{hholon}) and (\ref{sc}), we see that a
generalized Ginzburg-Landau (GL) equation can be written down for the holon
condensate $\psi _h({\bf r})$ as 
\begin{equation}
\alpha \psi _h+\beta |\psi _h|^2\psi _h+\frac 1{2m_h}\left( -i\nabla -{\bf A}%
^e-{\bf A}^s\right) ^2\psi _h=0~.  \label{gleqn}
\end{equation}
The current operator can be constructed in the usual manner as 
\begin{eqnarray}
{\bf J}({\bf r}) &=&-\frac i{2m_h}\left[ \psi _h^{\dagger }({\bf r})\nabla
\psi _h({\bf r})-\nabla \psi _h^{\dagger }({\bf r})\psi _h({\bf r})\right] 
\nonumber \\
&&-\frac{{\bf A}^e+{\bf A}^s}{m_h}\psi _h^{\dagger }({\bf r})\psi _h({\bf r}%
)~.  \label{current}
\end{eqnarray}
At $T=0$ and in the absence of an external magnetic field, all the spinons
are (RVB) paired. Then, we see from (\ref{As}) that ${\bf A}^s$ is trivial
and the phase of the superconducting condensate is robust. But {\em excited}
spinons can (and do) influence the holon condensate through the link field $%
{\bf A}^s({\bf r})$ in the above GL equation. In the following, we shall
investigate such effects systematically.

{\em Superconductivity, confinement, and $T_c$}: From the definition of the
order parameter (\ref{sc}), we infer that the presence of a holon condensate
is {\em not} sufficient for superconductivity to occur; {\em i.e.}, {\em %
phase coherence will not be realized in the system if the spinon vortices
unbind}, such that $<e^{i\Phi ^s({\bf r})}e^{-i\Phi ^s({\bf r}^{\prime })}>$
falls off exponentially. This leads us to define the transition temperature, 
$T_c$, as the temperature at which the spinon vortices bind, resulting in
the vanishing of $\Phi ^s$. To relate $T_c$ to the number of excited
spinons, let us first consider the energy it costs to create an isolated
spinon vortex. For a single spinon vortex centered around the origin, we
have $\displaystyle  
{\bf A}^s({\bf r})={\frac 12}{\frac{\hat{{\bf z}}\times {\bf r}}{r^2}}$, for
distances $r>>a_c$, the size of the vortex core. From (\ref{current}), we
see that the presence of a spinon creates a supercurrent in the holon
condensate through ${\bf A}^s$ [see also (\ref{supercurrent}) below]$:%
\displaystyle 
{\bf J}=-{\frac{\rho _h}{m_h}}{\bf A}^s$, as illustrated by Fig. 2(a).
Substituting the above in (\ref{hholon}), we get 
\begin{eqnarray}
E_v &=&-\int d^2{\bf r}\text{ }{\bf A}^s\cdot {\bf J}-\int d^2{\bf r}\text{{}%
}\rho _h{\frac{({\bf A}^s)^2}{2m_h}}  \nonumber \\
&=&{\frac{\rho _h}{2m_h}}\int d^2{\bf r}({\bf A}^s)^2 \\
&=&{\frac{\pi \rho _h}{4m_h}}\int dr{\frac 1r}\propto {\rm ln}{\frac L{a_c}}%
~,  \label{confinement}
\end{eqnarray}
where $L$ is the size of the sample. In general, the cost of creating more
than one spinon vortex can be obtained by substituting the expression for $%
{\bf A}^s({\bf r)}$, {\em viz.}, (\ref{As}) in (\ref{hholon}). It is easily
seen that the resulting expression describes a logarithmic attractive
(repulsive) interaction between vortex-antivortex (vortex) excitations, like
in the conventional Berezinskii-Kosterlitz-Thouless (BKT) transition.

From the above, we conclude that a single $S=1/2$ bosonic spinon excitation
is forbidden (owing to a logarithmically diverging energy) and all excited
spinons should be bound in vortex-antivortex pairs like in the low
temperature phase of a BKT system. Consequently, $\Phi ^s$ in (\ref{phis})
becomes trivial and phase coherence of (\ref{sc}) is realized. The ground
state is therefore a superconductor in which $S=1/2$ bosonic spinon
excitations are confined by a logarithmic potential. However, such a
confinement potential does {\em not} preclude the existence of fermionic
quasiparticle excitations. Elsewhere, we showed that $S=1/2$ nodal
quasiparticles can be created from the condensate as a composite
excitations. A calculation of the spectral function in the superconducting
RVB state shows the presence of quasiparticle peaks below $T_c$, consistent
with results from angle-resolved photoemission spectroscopy \cite{vnm_01}.
Furthermore, an $S=1$\ spin excitation can be constructed from a pair of
bosonic $S=1/2$ spinons, as shown in Fig. 2(b). The excitation energy $E_g$,
in this case, is finite and can be determined from the bosonic mean field
theory \cite{zy_99}. This excitation leads to a sharp feature in the
dynamical spin correlation function below $T_c$. Note that the vorticities
of the current vortices are not directly related to the spin polarization
directions, due to a gauge freedom to be discussed below.

The transition to the normal state occurs when the phase coherence of the
order parameter is destroyed; {\em i.e.}, $\Phi ^s$ in (\ref{sc}) is
disordered owing to the emergence of free spinon vortices and the
deconfinement of bosonic spinons marks the transition. It should be noted
that the temperature at which the unbinding of spinon vortex-antivortex
excitations takes place can be substantially lower than the conventional BKT
transition temperature ($\sim 1000$ K at optimal doping if ${\bf A}^s$ is
neglected \cite{nagaosa_90}). This is because the cores of the spinon
vortices begin to touch each other before the unbinding of
vortices-antivortices driven by entropy happens. In this dense limit, the
vortex-antivortex excitations can unbind because the energy cost is no
longer logarithmically divergent, and $T_c$ is the temperature at which the
average distance between spinons $l\simeq 2a_c$, where $a_c$ is the core
radius of a spinon vortex, as illustrated in Fig. 3. The latter can be
estimated by solving the GL equation for an isolated vortex \cite{later}.
Here, we merely point out that the spinons have a characteristic length
scale since their dynamics is governed by a uniform flux of $\delta \pi $
per plaquette. Recall that the spinons perceive the presence of holons as $%
\pi $ fluxoids. The gauge field $A_{ij}^h$ in (\ref{hspinon}) represents an
average flux of $\delta \pi $ (per plaquette) when the holons are condensed
and the spinons acquire a characteristic length scale, $a_c\sim a/\sqrt{\pi
\delta }$. Now, if $n_s^{ex}$ be the number of excited spinons, then, $l=2a/%
\sqrt{\pi n_s^{ex}}$. Therefore, we conclude that $T_c$ is determined by the
condition $n_s^{ex}=\delta $. The RVB mean field theory of (\ref{hspinon})
shows that this condition is satisfied at a temperature $T_c\sim E_g/4$ , $%
E_g$ being the characteristic spinon energy \cite{zy_99}, as plotted in Fig.
3, where the experimental data from YBCO\cite{bourges} are also shown for
comparison (taking $J=100$ meV). Such an estimate of $T_c$ based on the
``core touching'' mechanism agrees very well with the results from a
systematic renormalization group analysis of (\ref{hholon}) by M. Shaw {\em %
et al.} \cite{shaw_01}.

{\em Meissner effect and flux quantization}: We now consider the situation
when an external magnetic field is present at $T=0$. Writing $\psi _h({\bf r}%
)=\sqrt{\rho _h}e^{i\phi _h({\bf r})}$, we express the supercurrent given by
(\ref{current}) as 
\begin{equation}
{\bf J}={\frac{\rho _h}{m_h}}\left[ \nabla \phi _h-{\bf A}^e-{\bf A}%
^s\right] ~.  \label{supercurrent}
\end{equation}
Following the usual arguments for single valuedness of $\psi _h({\bf r})$,
we get 
\begin{equation}
{\frac{m_h}{\rho _h}}\oint_c{\bf J}({\bf r})\cdot d{\bf r}=2\pi n-\oint_cd%
{\bf r}\cdot \left( {\bf A}^e+{\bf A}^s\right) ~,
\end{equation}
where the integral is over a closed loop and $n$, an integer. Now suppose
that the integration is carried over a loop that is far away from the core
of the vortex. Then, ${\bf J}=0$ along the loop and we get 
\begin{equation}
\left( 2\pi n-\oint_cd{\bf r}\cdot {\bf A}^e\right) -\oint_cd{\bf r}\cdot 
{\bf A}^s=0~.  \label{fq1}
\end{equation}
When ${\bf A}^s=0$, we see that the magnetic flux is quantized at $2\pi n$
in units of $\hbar c/e$; {\em i.e.}, the minimal flux quantum in this case
is $hc/e\equiv \Phi _0$, as expected for a charge $e$ Bose system. However,
the presence of ${\bf A}^s$ changes the quantization condition radically.
Suppose there is one excited spinon trapped in the core of a magnetic
fluxoid [Fig. 4(a)]. Then, from (\ref{fq1}), we obtain the minimal flux
quantization condition, 
\begin{equation}
\oint_cd{\bf r}\cdot {\bf A}^e=\pm \pi ~,  \label{fq2}
\end{equation}
which is precisely the quantization condition of $\Phi _0/2$ in a
superconductor with $2e$ pairing. As the holons do not distinguish between
internal (fictitious) and external (magnetic) flux in (\ref{hholon}), they
still perceive a total flux quantized at $\Phi _0$ [see Fig. 4(a)], even
though the true magnetic flux quantum is $\Phi _0/2$.

{\em Stability of the $2e$ flux quantum}: When the magnetic flux quantum
inside the core is $\Phi _0$, there is no spinon trapped inside the core. On
the other hand, a bosonic spinon is trapped inside the core if the flux
quantum is $\Phi _0/2$. Therefore, we have to estimate the energy difference
between these two cases to determine which of these is energetically
favorable. The energy difference (per unit length) $\Delta \epsilon $
between $\Phi _0/2$ and $\Phi _0$ magnetic fluxes due to the magnetic field
and supercurrent is given by $\displaystyle  
\Delta \epsilon =-3\left( \frac{\Phi _0}{8\pi \lambda }\right) ^2\ln \kappa $
where $\lambda =\left( m_hc^2/4\pi e^2\delta \right) ^{1/2}$, $\kappa
=\lambda /a_c$, and $a_c\sim $ $1/\delta ^{1/2}$, as discussed earlier.
Then, $\displaystyle \Delta \epsilon =-\frac{3\pi }8t_h\delta \ln \kappa
\geq -t\delta $, which vanishes as $\delta \rightarrow 0$. So far, we have
ignored the cost of the vortex core. Since the $\Phi _0/2$ flux has a spinon
trapped inside the core, we need to add the energy cost of an excited spinon
to $\Delta \epsilon $ estimated above. Since the energy of an excited spinon
in the bulk (within the boson RVB theory\cite{zy_99}) is $E_s=E_g/2\sim
J\delta $, and since $t>J$, we expect the $\Phi _0/2$ flux quantum to be
favorable for all doping concentrations\cite{fn}. Furthermore, as will be
argued below, spinon (RVB) pairing inside the core is actually {\em improved}
over the bulk; {\em viz.}, the cost of creating a spinon inside the core is
smaller than the energy of an excited spinon in the bulk. This further
ensures the stability of the $\Phi _0/2$ flux quantum. This result has to be
contrasted with the slave boson RVB theories, where a $\Phi _0$ flux quantum
can be stabilized for small doping \cite{sachdev_92}, \cite{nagaosa_92}. In
a recent paper\cite{wynn_01}, Wynn {\em et al.}, carried out a careful
search for $\Phi _0$ fluxoids in the high temperature superconductors. Using
scanning SQUID and Hall probe studies, they looked for $\Phi _0$ fluxoids in
underdoped YBCO, but only observed $\Phi _0/2$ fluxoids \cite{wynn_01}. If
spin charge separation indeed occurs in the high temperature
superconductors, these experiments support our finding that only $\Phi _0/2$
fluxoids are favored energetically.

{\em Local moment inside a vortex core}: As discussed above, each magnetic
flux $hc/2e$ is associated with an unpaired spin $S=1/2$ trapped inside the
vortex core (Fig. 4), where the holon density is suppressed. Note that the
flux quantization condition (\ref{fq2}) is independent of the spin
polarization. While $\oint_cd{\bf r}\cdot {\bf A}^s=\pm \pi $ does depend on
the spinon polarization [{\em cf.}(\ref{As})], the resulting sign can always
be absorbed by the phase of the holon condensate ($\oint d{\bf r}\cdot
\nabla \phi _h=\pm 2\pi $), without changing $\oint_cd{\bf r}\cdot {\bf A}^e$
in (\ref{fq2}). To paraphrase, each magnetic flux can trap either an $%
S_z=+1/2$ or an $S_z=-1/2$ spinon {\em without} changing the minimal
quantization $hc/2e$. However, this two-fold degeneracy will be split by a
Zeeman coupling to the magnetic field. This leads to a very unique
consequence of the theory, discussed below.

The Zeeman splitting between the $S_z=\pm 1/2$ inside a vortex core can give
rise to a Schottky contribution to the low temperature specific heat. From
our analysis, we expect the number of magnetic moments [Fig. 4(b)] to be
directly proportional to the applied magnetic field H, at least for small
fields. The constant of proportionality can be estimated easily. Assuming a
lattice constant of $4A^{\circ }$ for the Cu-O planes, we find the number of
induced magnetic moments given by 
\[
n_{moment}\simeq \eta \times H~, 
\]
where $\eta =8\times 10^{-5}$ per Cu Tesla. Therefore, for a field of $6T$,
we estimate the number of induced magnetic moments to be $\sim 0.05\%$ per
Cu. Evidence for magnetic moments can be seen in the measurements of field
dependent specific heat at low temperatures. However, most of these
measurements are done on YBCO and the moments are invariably ascribed to
impurities associated with the chains. Here, we offer another explanation
based on our theory. The low temperature specific heat of the 123, 124 and
247 phases of YBCO has been measured by Sanchez {\em et al.} \cite
{sanchez_92}. They find that the observed specific heat in all these
compounds can be described satisfactorily if one were to include a magnetic
Schottky contribution whose amplitude ({\em i.e.}, the number of magnetic
moments) increases linearly with the field. For an applied field of $6T$,
they estimate the number of induced magnetic moments in all three phases of
YBCO to be $0.07-0.08\%$ per Cu atom. Similar results are observed in LSCO 
\cite{mason_93}, where the number of induced magnetic moments is also found
to increase linearly for small fields (up to $7T$) before saturating. Mason 
{\em et al.}, \cite{mason_93} estimate the number of induced magnetic
moments to be around $0.07\%$ per Cu at $6T$. These observations are in
agreement with our estimates for the number of induced magnetic moments, as
well as the field dependence. In another study of field dependent specific
heat in YBCO, Emerson {\em et al.}, \cite{emerson_94} determined the number
of induced magnetic moments to be $0.05-0.1\%$ per Cu in the same range of
magnetic fields. But they also find that the number of magnetic moments does
not vary with the applied field. However, as the authors point out, their
data is inherently different from the data of Sanchez {\em et al.} This
discrepancy needs to be resolved by future measurements. Recently, Moler 
{\em et al.}, \cite{moler_97} measured the field dependent specific heat of
YBCO and also observed the number of magnetic moments to be $\sim 0.1\%$ per
Cu.

Finally, we comment on the recent specific heat measurements in Zn doped
YBCO \cite{sisson}. These measurements show that the free magnetic moments
(induced by Zn substitution) that are present in the normal state do not
contribute to the specific heat below $T_c$. There is no Schottky anomaly
due to the magnetic moments for a wide range of Zn concentration in fields
up to $8T$. The authors explain this in terms of Kondo screening in the
superconducting phase. However, the present work may provide a different
explanation based on the confinement of bosonic spinons in the
superconducting state. Except for the free moments trapped at vortex cores,
no local moments related to $S=1/2$ bosonic spinons are allowed in the
superconducting bulk. This explains why the Schottky anomaly remains
unchanged as in the Zn-free case up to 1\% Zn concentration. 

{\em Enhanced AF correlations}: Thus far, we have seen that a spinon is
trapped inside the core of a vortex. Clearly, the amplitude of the holon
condensate $\rho _h$ vanishes at the trapped spinon position according to
the GL equation (\ref{gleqn}), consistent with the single occupancy
constraint. Since the spinon has a characteristic size $a_c$, we expect the
holon density to be reduced within such a core region. This, in turn,
enhances the RVB pairing inside the core. Consider the Hamiltonian (\ref
{hspinon}) for the spinons. At finite doping, the AF correlations are
suppressed by the vortex effect of the {\em holon} link field $A_{ij}^h$.
Now, if the holon density is reduced inside the core region, then there is a
reduction in the effect of $A_{ij}^h$ inside the core, and consequently, RVB
pairing or equivalently, the short range AF correlations are {\em enhanced}
compared to the bulk. In the extreme case of $A_{ij}^h\rightarrow 0$, when
there are no holon vortices, the correlations inside the core will mimic
those of the undoped compound. Thus, when a spinon is trapped near the
vortex core, there is a concomitant enhancement of AF correlations inside
the core region. This is a very important consequence of our theory and
should be contrasted with the results from slave-boson RVB theories. In the
latter, {\em both} the holon condensate and the RVB pairing are {\em %
suppressed} inside the core region. Consequently, AF correlations are absent
in the core region which is expected to be a normal Fermi liquid \cite
{sachdev_92}, \cite{nagaosa_92}. On the other hand, our considerations lead
to the conclusion that the core region is closer to the underdoped regime
with {\em enhanced} AF correlations.

Based on these arguments, we can obtain some simple estimates. Recall that
in the bulk, the characteristic length scale of a spinon, $a_c=a/\sqrt{\pi
\delta }$. Within the magnetic vortex core, due to the reduction of $\delta $%
, we expect the core size to increase, and the new core size, $l_c=\sqrt{%
\delta /\delta _c}a_c$, where $\delta _c$ denotes the average holon density
within the core. To be consistent with the condition that one holon is
expelled from the core region due to the trapping of an extra spinon, we
demand $\rho _c\times \pi l_c^2=\rho \times \pi l_c^2-1$, where $\rho
_c=\delta _c/a^2$ and $\rho =\delta /a^2$. We then get $\delta _c=\delta /2$%
. Thus, the holon density is reduced by half within the core and $l_c=\sqrt{2%
}a_c$. Since $E_g \varpropto \delta $ in the bulk, we conclude that the
characteristic spin excitation energy should be also approximately reduced
by half, {\em viz.}, $E_g^{core}\simeq E_g/2$. Recently, Lake {\em et al.},
reported the observation of AF correlations by neutron scattering inside the
vortex core \cite{lake_01}, where a field-induced sub-gap spin excitation
was found in LSCO at an energy scale $\sim 4.3$ meV, approximately half of
the characteristic spin energy scale in the bulk. The reduction of $%
E_g^{core}$ due to the reduced holon density may provide an explanation for
such an observation. Tunneling spectroscopy of the vortex cores shows that
the quasiparticle gap inside the core region is larger compared to the
magnitude of the bulk superconducting gap \cite{renner_98}. Again, this
result does not contradict our conclusion that the core region is closer to
the underdoped regime. These issues will be discussed in detail elsewhere 
\cite{later}.

To conclude, we presented an effective theory of the superconducting state
based on the bosonic representation of the $t-J$ model. The theory leads to
several interesting consequences. We first showed that the bosonic RVB
theory leads to phase vortices in the superconducting order parameter. The
phase vortices are excited spinons and $T_c$ is the temperature at which the
cores of these vortices begin to overlap. We showed how $hc/2e$ flux
quantization leads to the trapping of a spinon inside the vortex core and
argued that the cores exhibit enhanced antiferromagnetic correlations. Our
estimates for the core energy show that the bosonic RVB theory does not
allow for $hc/e$ flux quanta for any doping concentration. We also showed
that the trapping of a spinon inside the vortex core leads to observable
consequences for the low temperature specific heat. In a forthcoming
publication, we shall present a quantitative analysis of the structure of an
isolated vortex, based on the ideas outlined in this paper. We believe that
the approach presented in this paper can bridge the gap between a
microscopic model such as the $t-J$ Hamiltonian and effective theories of
quasiparticles in a $d-$wave superconductor coupled to fluctuating vortices 
\cite{qed3}. 

\acknowledgments 
We thank P. W. Anderson, Tao Li, N. P. Ong, Ming Shaw, and Yayu Wang for
very useful discussions. V.~N.~M. acknowledges partial support from NSF
Grant DMR 98-09483. 

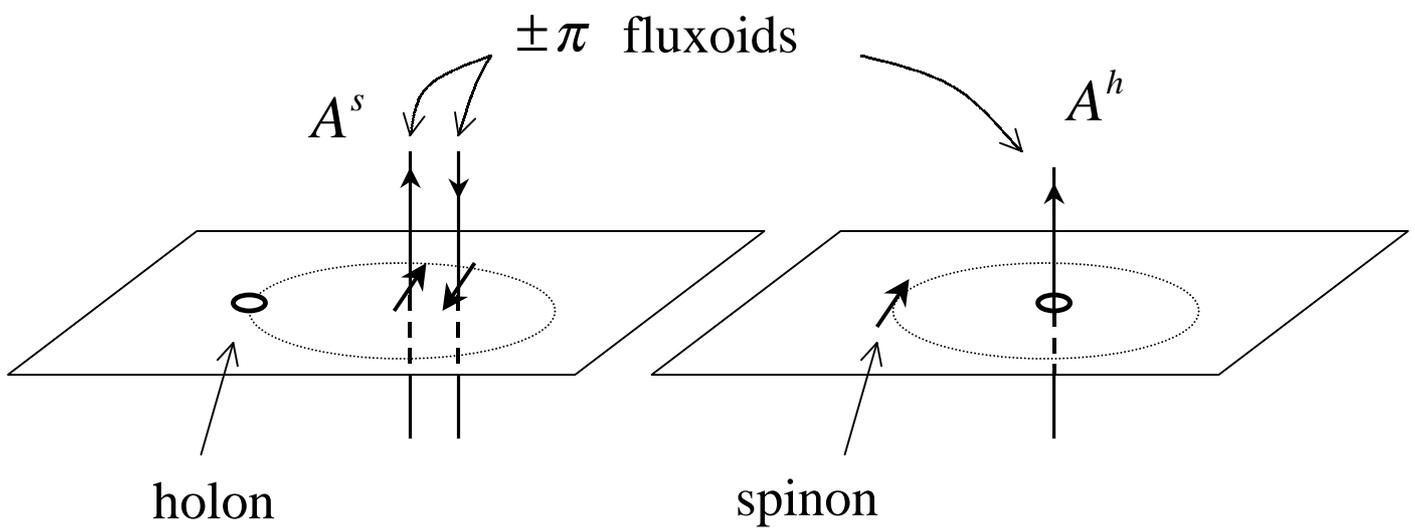
Fig. 1. Holons and spinons perceive each other as carrying fictitious $\pi$
flux tubes.

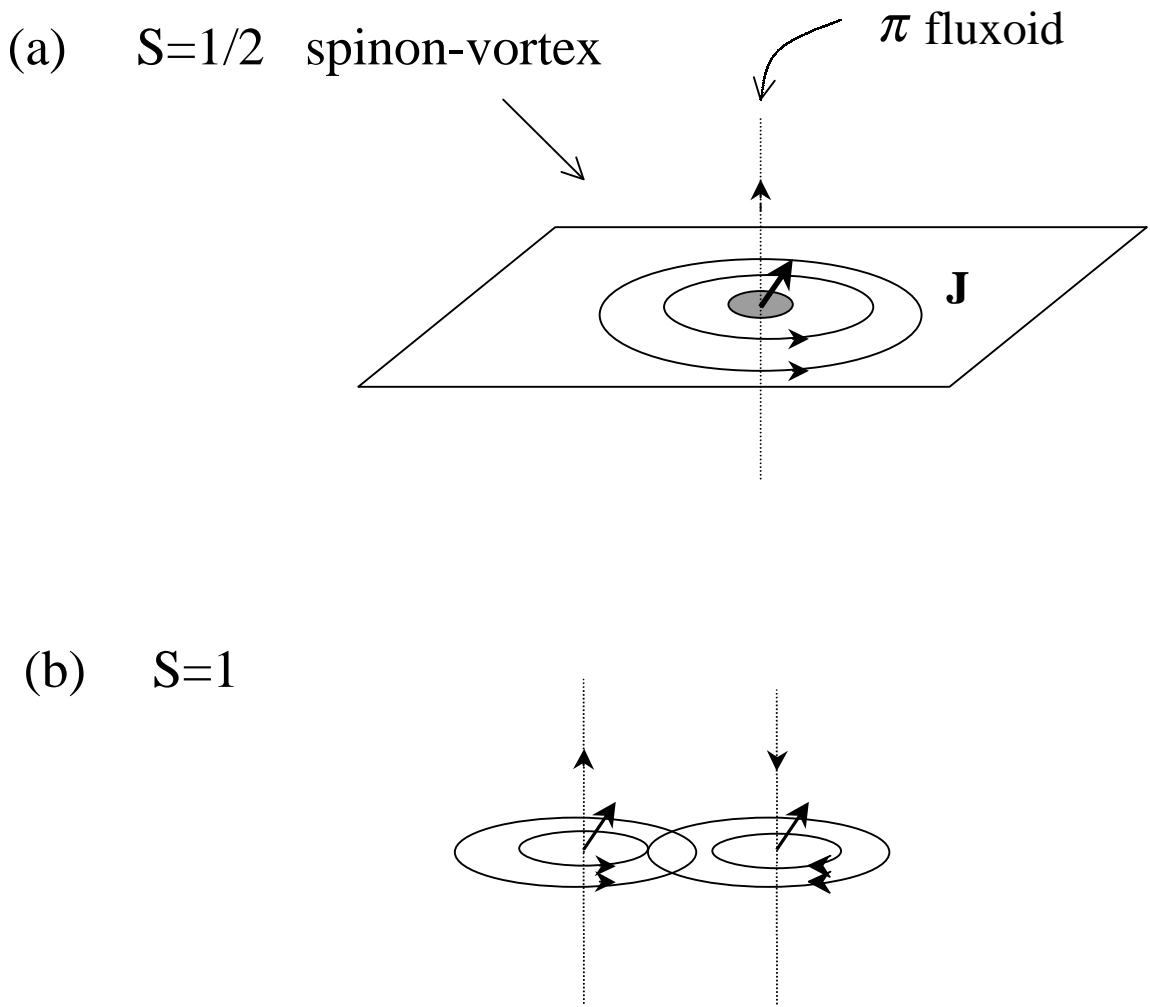
Fig. 2. (a) An $S=1/2$\ bosonic spinon carries a current vortex in the
superconducting phase; (b) An $S=1$ spin excitation is constructed from a
pair of confined $S=1/2$\ spinon vortices.

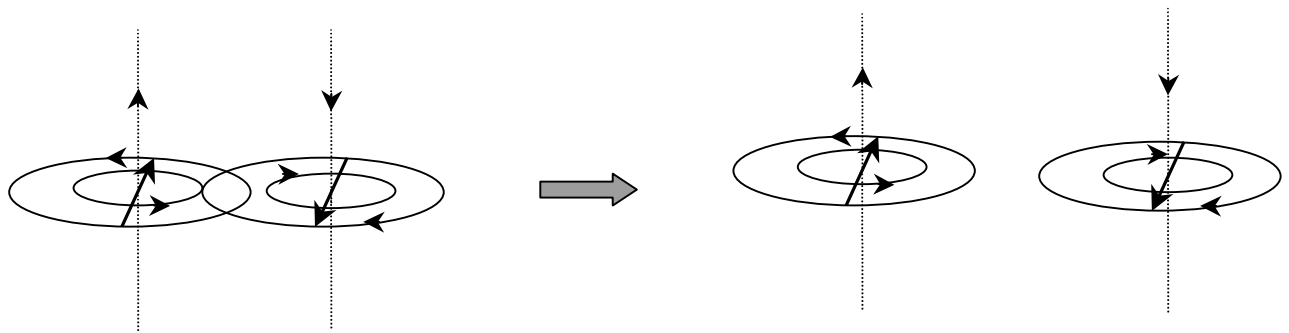
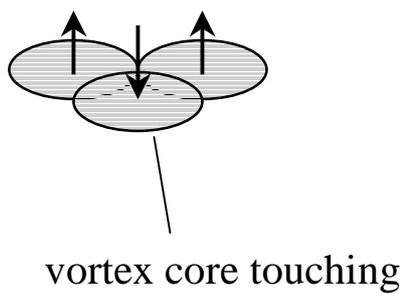
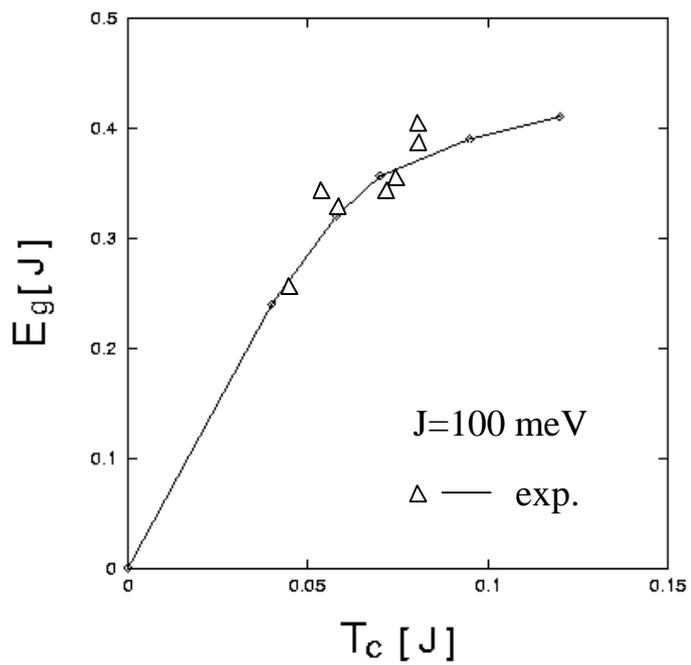
Fig. 3. Phase coherence disappears at $T_c$, as spinon vortices unbind due
to core touching; $T_c$\ is determined by the characteristic energy, $E_g$,
of the $S=1$ excitation. The experimental data are from Ref. \cite{bourges}.

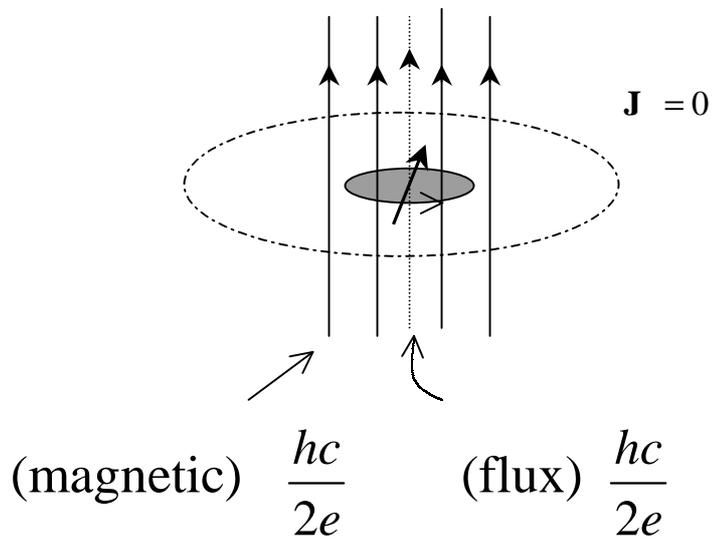
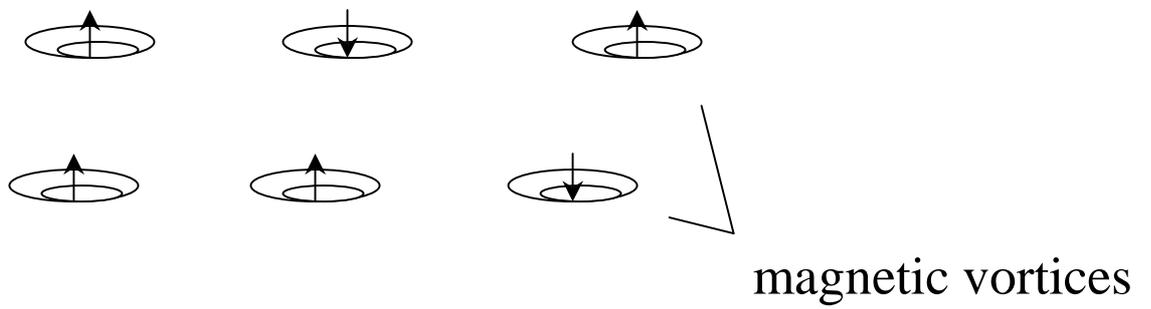
Fig. 4. Flux quantization occurs at $hc/2e$, with a bosonic $S=1/2$ spinon
trapped inside the core.


\begin{references}
\bibitem{fn}  At optimal doping, using $\lambda \sim 1500$ 
$A^{\circ }$, we estimate the energy difference between the $\Phi_0$ and
$\Phi_0/2$ flux quanta to be $\sim 800$ meV, much larger than $E_s\sim
20$ meV ($E_g\sim 41$ meV\cite{bourges}).
\end{references}
\end{document}